\documentclass[sigconf]{acmart}
\AtBeginDocument{%
  \providecommand\BibTeX{{%
    \normalfont B\kern-0.5em{\scshape i\kern-0.25em b}\kern-0.8em\TeX}}}

\copyrightyear{2021} 
\acmYear{2021} 
\setcopyright{acmlicensed}
\acmConference[AIES '21]{Proceedings of the 2021AAAI/ACM Conference on AI, Ethics, and Society}{May 19--21, 2021}{VirtualEvent, USA}
\acmBooktitle{Proceedings of the 2021 AAAI/ACM Conference on AI, Ethics,and Society (AIES '21), May 19--21, 2021, Virtual Event, USA}
\acmPrice{15.00}\acmDOI{10.1145/3461702.3462519}
\acmISBN{978-1-4503-8473-5/21/05}

\begin{document}
\fancyhead{}

\title{Emergent Unfairness in Algorithmic Fairness-Accuracy Trade-Off Research}

\author{A. Feder Cooper}
\email{afc78@cornell.edu}
\affiliation{%
  \institution{Cornell University}
  \department{Department of Computer Science}
  \city{Ithaca}
  \state{NY}
  \country{USA}
  \postcode{14850}
}

\author{Ellen Abrams}
\email{ema85@cornell.edu}
\affiliation{%
  \institution{Cornell University}
  \department{Society for the Humanities}
  \city{Ithaca}
  \state{NY}
  \country{USA}
  \postcode{14850}
}

\renewcommand{\shortauthors}{Cooper and Abrams}

\begin{abstract}
  Across machine learning (ML) sub-disciplines, researchers make explicit mathematical assumptions in order to facilitate proof-writing. We note that, specifically in the area of fairness-accuracy trade-off optimization scholarship, similar attention is not paid to the \emph{normative assumptions} that ground this approach. Such assumptions presume that 1) accuracy and fairness  are in inherent opposition to one another, 2) strict notions of mathematical equality can adequately  model fairness, 3) it is possible to measure the accuracy and fairness of decisions independent from historical context, and 4) collecting more data on marginalized individuals is a reasonable solution to mitigate the effects of the trade-off. We argue that such assumptions, which are often left implicit and unexamined, lead to inconsistent conclusions:  While the intended goal of this work may be to improve the fairness of machine learning models, these unexamined, implicit assumptions can in fact result in emergent unfairness. We conclude by suggesting a concrete path forward toward a potential resolution.
\end{abstract}






\maketitle

\section{Introduction} \label{sec:intro}
Optimization is a problem formulation technique that lies at the core of multiple engineering domains.
Given some fixed or limited resource, we can model its usage to effectively solve a problem. The optimal solution is the one that either minimizes some cost function or maximizes some utility function---functions that measure how well the model performs on a particular objective. Often there is more than one objective to satisfy simultaneously, and those objectives can be in tension with one another. In this case, it is possible to pose this problem as optimizing a trade-off \cite{yang2010optimization}.

\begin{figure}[ht]
    \centering
    \includegraphics[width=0.8\linewidth]{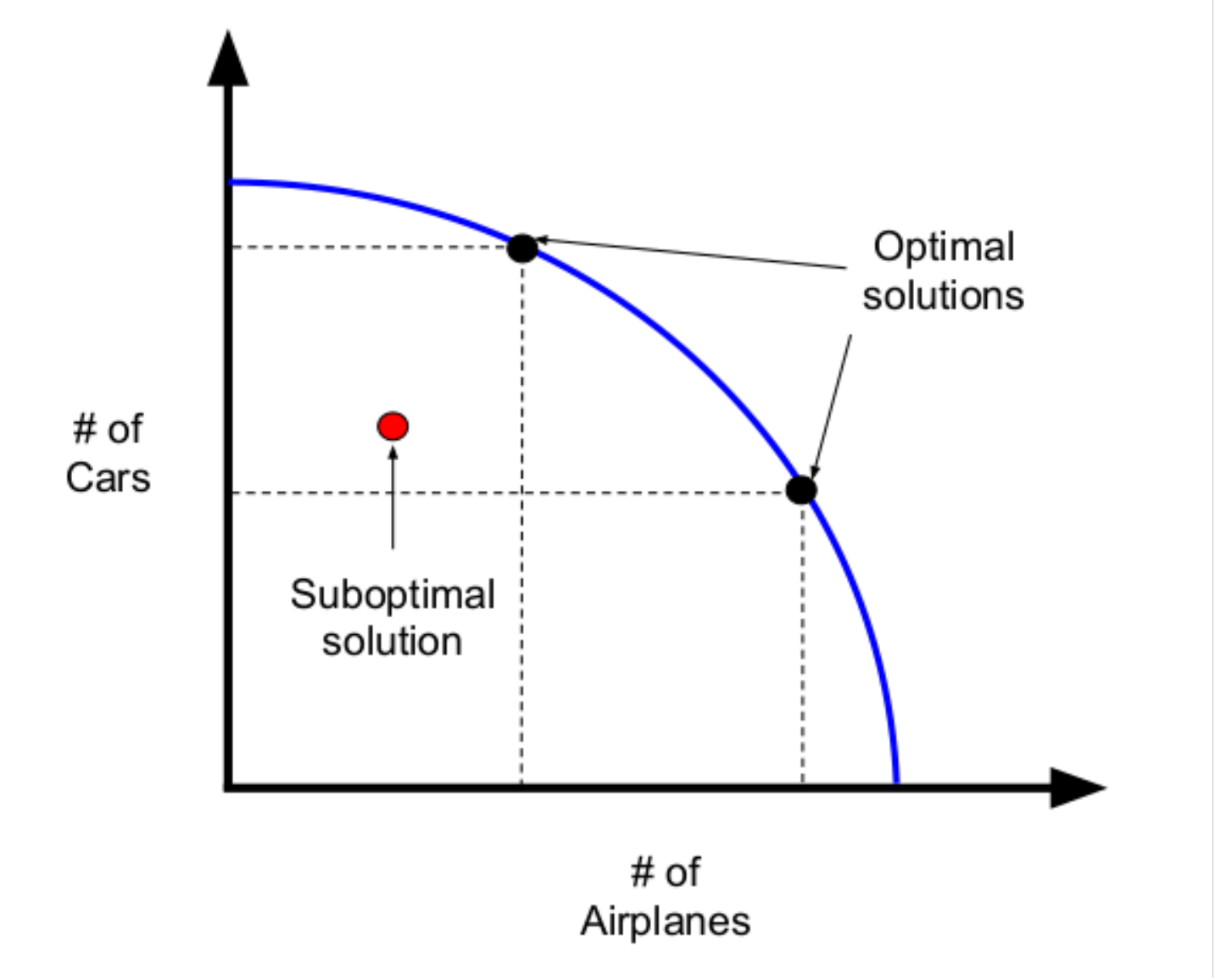}
    \caption{Illustrating a trade-off. Given a fixed amount of steel, a company can build a combination of cars and planes. To optimally utilize the steel for maximizing profit, it can manufacture any point on the blue curve. Any combination above the curve is not possible because there is insufficient steel; any below is suboptimal because the same amount of steel could either produce more cars or more planes.}
    \label{fig:optimization}
\end{figure}

For an intuitive example, consider a company that has a fixed amount of steel, which it can use to build cars and planes, which it then sells to earn a profit. The company has to decide how to allocate the steel to maximize that profit and can formulate the decision as an optimization problem. The blue curve in Figure \ref{fig:optimization} models possible ways to do this optimally; picking a specific point on the curve corresponds to the company's choice for how to balance the trade-off between how many cars and how many planes to produce.

Optimization science has informed much of the last decade's spectacular spate of statistical machine learning (ML) publications. It is at the core of how many ML algorithms learn. For example, learned classifiers use training data examples to fit a curve that optimizes for both classifying those examples correctly and generalizing to new, unclassified examples \cite{bishop1995mlbook, hastie2009statistical}. This process of automating classification decisions has in the past been framed as optimal in another sense; automation brought with it the hope of rooting out the human (suboptimal) whims of decisionmaking---of eliminating the ugliest human biases, such as sexism and racism, that afflict high-impact decision processes. Yet, as has been well-documented, this hope cannot be passively realized merely by substituting humans with automated decision agents. Issues with biased data and biased model selection processes can in the worst case magnify, rather than replace, human biases \cite{abebe2020social, forde2021actionable, selbst2019abstraction}.

In response, there has been a widespread push to actively engineer \emph{algorithmic fairness}. This has become remarkably urgent as automated decision systems are being deployed in domains that have a significant impact on the quality of human life---in deciding judicial bail conditions, loan allocation, college admissions, hiring outcomes, COVID vaccine distribution, etc., \cite{monahan2016risk, barocas2014datas, ajunwa2020hiring, angwin2016pp}. Models no longer just need to be correct; they also need to be fair.

It has become common to position accuracy in opposition to fairness and to formalize a mathematical trade-off between the two. For example, in the context of criminal justice and bail decisions, the accuracy of decisions has been framed as how to best ``maximize public safety," in contrast to satisfying ``formal fairness constraints" that aim to reduce racial disparities in decision outcomes \cite{corbettdavies2017cost}. This kind of problem formulation is the norm in a growing area of research, in which the trade-off between fairness and accuracy in ML is considered ``inherent" or ``unavoidable." Prior work suggests various ways of implementing the trade-off: At best, under particular conditions, the tension between the two can be dissolved to achieve both; at worst, fairness is sacrificed in favor of accuracy, while the remaining cases fall somewhere in the middle~\cite{dutta2020tradeoff, chen2018active, bakker2019active, menon2018cost, noriegacampero2019active}.

\subsection{Our Contribution}
Our work looks both at and beyond the fairness-accuracy trade-off, drawing from prior critiques of algorithmic fairness and studies of sociotechnical systems. We examine the \emph{choice}---and our work here will show that it is a choice, not a requirement---to model assumptions that cast fairness in direct opposition to accuracy. Regardless of the particulars of specific implementations, this framing does not just involve math, but also implicates normative concerns regarding how to value fairness and accuracy both independently and in relation to each other \cite{flanagan2014values, friedman2019valuesbook}. 

Our contribution is to extract and explore patterns of these concerns across trade-off scholarship that arise at three different stages: the initial modeling assumption to treat accuracy and fairness together in an optimization problem, the move from abstract framing to concrete problem formulation, and the ``optimal solutions" that result from those formulations. More specifically, we examine how the choice to operationalize the relationship between fairness and accuracy using the language of optimization inherently puts the two in conflict, rather than leaving open the legitimate possibility for them to actually be in accord. We discuss how this choice fails to take full account of social criteria when drawing the boundaries of a problem \cite{selbst2019abstraction}, and relates to broader trends of techno-``solutionism," in which math is (mistakenly) bestowed special authority to ``solve" social problems \cite{abebe2020social}.

Beyond  these overarching framing assumptions, there are other underlying, unexamined normative assumptions (not just explicit mathematical ones) that take root in how the trade-off is formalized: That strict notions of mathematical equality can model fairness, that it is possible to measure the accuracy and fairness of decisions independent from historical context, and that collecting more data on marginalized individuals---a practice called \emph{active fairness}---is a reasonable solution to mitigate the effects of the trade-off.

If we take the time to clarify these implicit assumptions, we note that the conclusions that follow can actually perpetuate unfairness: The mathematical proofs may be sound---a particular choice of fairness metric may even be optimized---but the implicit normative assumptions and accompanying broader normative results suggest that these methods will not ensure fairer outcomes in practical applications. In summary, we argue that
\begin{itemize}
    \item Using the language of optimization situates fairness and accuracy in intrinsic opposition, generally privileging the latter over the former and necessarily foreclosing the possibility of examining the ways they can instead reinforce each other (Section \ref{sec:sociotechnical}).
    \item Underlying mathematical assumptions bring unexamined normative dimensions, which can actually result in emergent unfairness (Section \ref{sec:assumptions}).
    \item In light of these observations, algorithmic fairness researchers will conduct more robust research if they first clarify their normative assumptions (Section \ref{sec:resolution}).
    \item Due to the extent of emergent unfairness from conceiving of fairness and accuracy in trade-off, it is worth revisiting this formulation altogether (Section \ref{sec:conclusion}).
\end{itemize}

\section{The Fairness-Accuracy Trade-Off} \label{sec:preliminaries}

We begin by providing the background necessary for understanding the problem formulation of the fairness-accuracy trade-off. Before clarifying what the trade-off actually characterizes, we address each component in turn, summarizing common quantifiable metrics that ML researchers map to the values of ``accuracy" and ``fairness."

\subsection{Accuracy Metrics} \label{sec:accuracy}

In brief, \emph{accuracy} measures how often a ML model correctly predicts or infers a decision outcome after training. So, to understand accuracy, we need to understand how ML models are trained. For the classification problems that dominate much of the fairness literature, training the model usually entails fitting a curve to a set of training data points for which classification labels are already known. After training, when supplied with a previously-unseen data point, the model can infer the classification label for that data point. For example, in building a model that infers whether or not to grant an applicant a loan, the curve-fitting training process occurs with past data concerning loan-granting decisions; inference corresponds to the model receiving a new loan applicant's data and classifying whether that applicant should receive a loan or not---a decision ultimately corresponding to whether or not the loan-granting institution believes the applicant will repay or default on the loan.

A model's accuracy tends to be measured during a validation process in between training and inference. Rather than using all labeled data for training, researchers reserve a portion to validate how well the trained model classifies unseen data points with known classification labels. In other words, accuracy is often a measure of \emph{label alignment}: It is the percentage of correctly classified validation data points, where correctness is determined by whether the model's classification decision matches the known label. There are other metrics that researchers use, such as Chernoff information \cite{dutta2020tradeoff}; however, label alignment is a popular accuracy metric, in part due to its simplicity.

This simplicity can be misleading, both in terms of what the math is actually measuring and the normative implications of that measurement. The broader algorithmic fairness community (and corresponding community of critics) has paid ample attention to this issue in relation to fairness \cite{binns2018phil, selbst2019abstraction, powles2018bias, abdurahman2019response}; however, in fairness-accuracy trade-off literature, where accuracy is also explicitly centered as a value, parallel analyses of accuracy have been relatively sparse.  We examine this in Section \ref{sec:assumptions}; for now, we emphasize that something as simple as a percentage metric can raise normative concerns \cite{forde2021actionable}.

One can see this from work in the broader ML community, in which accuracy issues often get cast as a problem of \emph{label bias}: The classification labels in the training and validation data can be incorrect, in terms of some abstract notion of ``ground truth."\footnote{The implications of ``true" classification, including the simplifying assumptions that inform such classification, are out of scope for our purposes. We refer the reader to the rich literatures in sociology and science \& technology studies on categorization and classification, notably~\cite{velocci2021diss, bowker1999sorting}.} As an innocuous example, consider a labeled image dataset of dogs and cats. The individual that labeled the dataset incorrectly (though perhaps understandably) mis-labeled Pomeranians as cats. This mis-labeling in turn leads the learned ML model to mistakenly identify Pomeranians as cats. 

The results of mis-labeling can be devastating for applications that impact human lives. Consider again an automated decision system that grants and denies loan applications. In the US, systemic racism against Black loan applicants, specifically manifested in the practice of redlining, has entailed denying loans to qualified Black applicants. These applicants, in terms of ``ground truth" should have been granted loans, but were instead intentionally marked as likely defaulters. As with the example above, this mis-labeling would lead to inaccurate classification. The models trained on that data would mistakenly identify Black non-defaulters as defaulters---a mis-classification that could be used to wrongfully deny a loan.\footnote{We have simplified this example, which also involves using zip code as a proxy for race, to make our point that mis-labeled data impacts the accuracy of trained models.} Unlike the former example, this one involves intentional mis-labeling; however, it is worth noting that even if the mis-labeling were unintentional, it would have the same effect on classification decisions. Regardless of intention, the impact would be the same. Either way, the learned model would systematically, incorrectly classify Black individuals as defaulters.

\subsection{Fairness Metrics} \label{sec:fairness}

Algorithmic \emph{fairness} has dozens of mathematical definitions that can inform optimization problem formulation.\footnote{Similar to our discussion of accuracy and classification, our work will not focus on the normative limits of defining fairness mathematically, such as the challenges of formulating fairness problems that account for intersectional protected identities \cite{selbst2019abstraction, chen2018active, dwork2018decoupled, buolamwini2018gender, kroll2017accountability, hoffman2019fairness} and how many fairness definitions reflect \emph{what we ought to believe} instead of (the arguably more useful, from a policy perspective) \emph{what we ought to do} \cite{hellman2019fairness}.} All definitions, regardless of the specifics, involve some treatment of \emph{protected attributes}, such as race and sex, along which decision outcomes can be evaluated for ``fair" treatment. Broadly speaking, there are two families of fairness metrics: Those that measure individual-focused fairness and those that evaluate it in terms of groups defined by protected attributes.

Individual fairness, as the name suggests, centers analyzing automated decisions in terms of the individual \cite{dwork2012fairness, joseph2016rawls, bakker2019active}. %For example, counterfactual fairness deems a decision to be fair toward an individual if it is the same regardless of the individual's membership in a particular demographic group \cite{kusner2017counterfactual}.
In contrast, group fairness centers demographic group membership and typically aims to ensure that membership in a protected class does not correlate with decision outcomes \cite{dwork2018decoupled, chen2018active}. A particularly popular metric is \citet{hardt2016equality}'s formulation of \emph{equality of opportunity}, which in essence only requires that there is no discrimination based on demographics for those assigned the positive classification. For the example of granting loans, paying back the loan is the positive class and defaulting is the negative class. Equality of opportunity corresponds to making sure that the rate of mistakenly classifying non-defaulters as defaulters is similar across demographic groups. We discuss this further in Section \ref{sec:assumptions}; for now, it is important to note that \emph{this formulation is inextricably tied to accuracy}. Training a model to reduce mistaken negative classification directly depends on training data that contains negative labels. Those labels, however, may not align with what is correct in terms of ``ground truth"---e.g., Black individuals erroneously marked as defaulters.

Lastly, it is also worth noting that multiple different mathematical notions of fairness cannot be satisfied simultaneously. This incompatibility has been formalized as impossibility results \cite{kleinberg2017impossibility,chouldechova2017impossible} and has placed a more significant emphasis on how computer scientists choose which fairness metric to study \cite{friedler2016impossibility}. While these findings are extremely well-cited, they are not surprising when considering fairness beyond its mathematical definition as a metric. In a pluralistic world, values like fairness depend on the time and place in which they are defined; outside of math, different, incompatible definitions can hold simultaneously, depending on context~\cite{berlin2013crooked}.

\subsection{The ``Inherent" Trade-Off} \label{sec:tradeoff}

Algorithmic fairness tends to pose accuracy and fairness in an ``inherent" or ``unavoidable" trade-off: An increase in fairness necessarily comes with a decrease in accuracy; increasing accuracy necessarily decreases fairness \cite{chen2018active, menon2018cost, bakker2019active, corbettdavies2017cost, dwork2018decoupled, sabato2020bounding, zhao2019inherent}. 

How did this trade-off problem formulation come about? Much of the literature that engages this trade-off does so empirically; that is, the authors perform or cite experiments that convey the intuition that a trade-off exists, so they decide that a mathematical trade-off is an appropriate way to model the problem. However, most do not characterize or quantify the trade-off theoretically. The work that has attempted theoretical treatment suggests that---at least in theory---the existence of the fairness-accuracy trade-off can be less rigidly described \cite{dutta2020tradeoff, wick2019revisited}. Nevertheless, the general practice in the field is to tacitly accept the trade-off as fact, regardless of the particular fairness and accuracy metrics under consideration.

Computer scientists further observe that the ramifications of this trade-off, particularly in high-stakes domains, can be significant. As a result, they sometimes wade into the murkiness of \emph{how} to optimize the trade-off implementation in specific ``sensitive" applications. For example, several computer scientists have noted that in areas like healthcare, trade-off implementations should favor accuracy, as privileging fairness can have ``devastating" consequences, such as missing cancer diagnoses at higher rates \cite{chen2018active, srivastava2019perceptions}.

Other researchers have posited \emph{why} this trade-off exists in the first place. For example, one popular explanation comes from \citet{friedler2016impossibility}. The authors of this paper reason that there is an ``abstract construct space" that represents the features we actually want to measure but cannot observe (e.g., intelligence). Instead, we see features in the ``observed space" of the actual world (e.g., SAT score), and there is a mapping from features in the construct space to features in the observed space (e.g., SAT score is the mapped feature in the observed space, standing in place for intelligence in the construct space). According to \citet{friedler2016impossibility}, the trade-off between classification accuracy and fairness exists in the real world due to ``noisier mappings" for less privileged groups from the construct space to the observed space. They contend that this noise comes from historic differences, particularly in opportunity and representation, which makes positive and negative data points less distinguishable (in comparison to privileged groups) for the learned classifier. It is this decrease in separability that leads to less fair classification for less privileged groups. In the example of SATs, this means that the scores are less reliable in terms of conveying information about intelligence for underprivileged groups. While this posited explanation may seem reasonable, work that engages with it rarely (if ever) supports the explanation with data, as it is usually not the specific fairness problem under mathematical consideration~\cite{dutta2020tradeoff}.

\section{Applying a Sociotechnical Lens}\label{sec:sociotechnical}
In the field of Science and Technology Studies, the term ``sociotechnical" is used to signal the impossibility of understanding the social world separately from the technological, and vice versa~\cite{bijker1987sociotechnical, hughes1993power, hecht2002nuclear, wajcman2004feminism}. This means that, among other things, technologies like algorithms are never separate from the systems, institutions, ideologies, cultures, and societies in which they are embedded. It also means that the choice to formulate algorithmic fairness as an optimization problem produces a particular kind of knowledge about fairness that cannot be detached from its broader social context.
%In the field of Science and Technology Studies, the term ``sociotechnical" is used to indicate the presence and interaction of both social and technical components in systems like the electric power grid \cite{hughes1993power}, uranium mining practices \cite{hecht2002nuclear}, Wikipedia organization \cite{geiger2017wiki}, and algorithmic decision-making \cite{selbst2019abstraction}. 
In our critiques of emergent unfairness in Section \ref{sec:assumptions}, we explicitly examine the fairness-accuracy trade-off model as a sociotechnical system.

We are not the first to examine algorithmic fairness using the concept of a sociotechnical system. For example, 
\citet{selbst2019abstraction} use the concept of sociotechnical systems to draw attention to five pitfalls that imperil well-meaning data scientific approaches to fairness. 
%Overall, the authors argue that the use of ``bedrock" computer science concepts in fairness research can lead to problematic outcomes. By focusing on one such bedrock concept---optimization---that characterizes a newly popular approach to algorithmic fairness research, we offer additional specificity and insight to existing critiques.
One that they identify is called the ``Framing Trap," which the authors describe as the ``failure to model the entire system over which a social criterion, such as fairness, will be enforced." A sociotechnical lens, they argue, might suggest new ways for researchers to draw the boundaries of fairness problems to include social relations and dynamics that may have otherwise been excluded. 

Defining fairness and accuracy in trade-off both exemplifies falling into the Framing Trap and presents an additional set of nuanced consequences. For example, the language of defining trade-offs along a curve (Figure \ref{fig:optimization}) necessarily requires a give-and-take relationship between the factors under consideration, as those factors are cast in opposition to one another. In the fairness-accuracy trade-off, fairness and accuracy are framed as inherently competing goals, in which we must give up some of one in order to gain some of the other (even if ``some" cannot be quantified definitively). Based on this framing, it is consistent that much of the literature uses language that describes the ``cost of fairness" \cite{chen2018active, menon2018cost, corbettdavies2017cost, dutta2020tradeoff}, depending on where on the trade-off optimization curve a particular implementation lies. This cost, however, can be described as cutting both ways: It is similarly reasonable to talk about the ``cost of accuracy." Yet, with few exceptions \cite{corbettdavies2017cost}, the literature in this area chooses not to discuss costs in this way. This decision is perhaps due to the tendency within the field of ML more broadly to privilege accuracy during model design; nonetheless, this framing shifts the burden of defensibility to fairness, in the sense that it implies that fairness' ``costs" require justification.

The particulars of the trade-off choice present additional complications. Beyond failing ``to model the entire system", the fairness-accuracy trade-off formulation also forecloses the very reasonable possibility that accuracy is generally in accord with fairness (unless one specifies particular conditions for which it is possible to demonstrate that the trade-off does not exist \cite{dutta2020tradeoff, wick2019revisited}). In other words, a trade-off model conceals the idea that the accurate thing to do could be complementary with the fair thing to do. For example, as \citet{hellman2019fairness} notes, it is possible to view accuracy and fairness as complementary values, where the former reflects what one ``ought to believe" and the latter reflects what one ``ought to do." In Section \ref{sec:assumptions}, we expand on how fairness and accuracy can be considered in accord \cite{friedman2019valuesbook, flanagan2008values}. Additionally, we suggest that the boundaries of the trade-off problem need to be redrawn to account for social and technical considerations \emph{in historical context}. A suitable frame for fairness research can neither be blind to past historical context nor ignore the future. 

Lastly, our analysis demonstrates some of the challenges of using trade-off and optimization tools in algorithmic fairness research. Borrowing tools from adjacent fields in computer science not only affects the results of fairness research, but it also helps to characterize the disciplinary status of fairness research itself. In this case, the operationalization of fairness as a mathematical problem helps situate questions of fairness within the realm of ``science," thus conferring a particular legitimacy that science connotes \cite{gieren1983boundary,porter1995trust}. Situating the fairness question as a scientific question and esconcing it in the language of trade-offs and optimization suggests that it is reasonable to try to solve for an ``optimal," best answer. Framing the problem as a trade-off problem to be ``solved" using math falls under the highly-critiqued practice of technological ``solutionism" \cite{selbst2019abstraction, abebe2020social}---the notion that technology is uniquely capable of solving social problems. This tendency toward solutionism grants special legitimacy to algorithmic fairness research, legitimacy absent in other fields that have tackled but have not ``solved" the fairness problem from a social perspective. We contend that it is unlikely that the same ideas used to solve problems of steel allocation (Figure \ref{fig:optimization}) will transfer without issue to questions of fair hiring practices. We suggest that attending to the sociotechnical context of each situation may help prevent the emergent unfairness we identify in the following sections.

\section{Emergent Unfairness}\label{sec:assumptions}

Choosing to convey fairness and accuracy as a trade-off is a mathematical modeling assumption. As we discussed in Section \ref{sec:preliminaries}, the authors have observed a pattern in their empirical results concerning accuracy and fairness, and deem a trade-off to be a useful way to formulate the mathematical problem of characterizing the relationship between the two. As we suggest above, fairness and accuracy metrics have normative dimensions; so, too, does the modeling assumption that poses them in trade-off. 

There are also numerous, other mathematical assumptions, each which carry their own implicit normative dimensions. We observe that, based on these implicit assumptions, fairness-accuracy trade-off scholarship is plagued with gaps and oversights. These issues can lead to conclusions that \emph{actually perpetuate unfairness}. There are dozens of examples of particular assumptions specific to each paper in fairness-accuracy trade-off scholarship. It is not possible to be exhaustive regarding each mathematical assumption's corresponding normative assumptions. Instead, we isolate three patterns of implicit, unexamined assumptions in the discipline, and the emergent \emph{un}fairness that can result: Unfairness from assuming 1) strict notions of equality can substitute for fairness, 2) historical context is irrelevant when formulating the trade-off, and 3) that collecting more data on marginalized groups is a reasonable mechanism for alleviating the trade-off.

\subsection{Unfairness from Assuming Fairness = Equality}

One assumption prevalent in fairness-accuracy trade-off literature concerns how different papers choose to measure fairness. Most of the work in this subfield relies on parity-based definitions. Algorithmic fairness definitions like this effectively make the modeling assumption to represent ``fairness" as ``equality."\footnote{While such equality-based notions of fairness dominate the literature more generally, not just concerning the fairness-accuracy trade-off, there are a growing number of exceptions. For example, some recent work frames fairness in terms of Rawlsian social welfare~\cite{rawls1971theory,heidari2018rawls, joseph2016rawls}.} To be clear, we mean ``equality" in the strict sense of the values of metrics being as equal as possible, by minimizing some form of measured inequality. This kind of strict equality can stand in for ``fairness" in terms of what is actually being modeled; fairness is being framed as a problem of strict equality. This is easily discernible in the popular equality of opportunity metric (Section \ref{sec:preliminaries}), used in numerous trade-off papers \cite{chen2018active, dutta2020tradeoff, bakker2019active, noriegacampero2019active}, which tries to minimize discrepancies in false negative classification decisions among different demographic groups; it literally tries to make those rates as equal as possible.

However, what is fair and what is equal are not always the same thing, and framing them as equivalent can actually lead to unfair outcomes \cite{kasy2021equal}. For example, when addressing historic or systemic inequity, it can be necessary to take corrective or reparative action for some demographic groups in order to create the conditions of more-equal footing. Such actions necessarily diverge from equality in the strict mathematical sense, so strictly parity-based fairness metrics cannot capture this kind of nuance.

The ongoing debate in the United States around the fairness of affirmative action policy can help illustrate this distinction, as well as the complications that arise when defining fairness as equality. In brief, affirmative action is a social policy aimed at increasing the representation of historically marginalized groups in university student and workforce populations; it attempts to implement a fairer playing field by providing individuals from marginalized backgrounds with the chance to have the same opportunities as those from more privileged backgrounds. 

While affirmative action has existed as official policy in the US for decades \cite{kennedy1961affirmative}, it is extremely contentious.
Many Americans, who do not feel personally responsible for systemic discrimination against BIPOC\footnote{An acronym for ``Black, Indigenous, and People of Color," used particularly in the US and Canada.} populations, feel that affirmative action puts them at a disadvantage. They claim that the policy is unfair, and in fact is responsible for ``reverse discrimination" \cite{budryk2020ut, newkirk2017white, pham2018harvard}. This belief comes in part from the idea that affirmative action does not lead to ``equal" comparisons in the strictest sense---comparing SAT scores or GPAs point for point. Instead, in the language of \citet{friedler2016impossibility}, one could say that affirmative action attempts to repair or normalize for the ``noisy mappings" that these scores convey for unprivileged populations in order to promote fairer outcomes.\footnote{It is also interesting to note that this controversy has found its way into the language of algorithmic fairness literature. \citet{dwork2012fairness, dwork2018decoupled} use the term ``fair affirmative action" in their work; they seem to be attempting to distinguish their notion from some imagined, alternative, unfair variant. This term is arguably redundant, since affirmative action is fundamentally about trying to promote fairer outcomes, even if that notion of fairness does not align with strict-equality-based notions in algorithmic fairness.} 

In short, the goal of affirmative action illustrates how notions of fairness and equality can diverge. The policy's existence is predicated on the notion that strictly equal treatment, without attending to past inequity, can potentially perpetuate unfairness.

\subsection{Unfairness from Assuming the Irrelevance of Context}

Fundamentally, the issue with the assumption that strict notions of equality can stand in for fairness has to do with how the assumption treats---or rather discounts---context. Fairness metrics like equality of opportunity are only able to evaluate the local, immediate decision under consideration. As discussed above using the example of affirmative action, this type of equality cannot accommodate reparative interventions that attempt to correct for past inequity. This similarly implicates issues with how we measure accuracy, since such metrics measure the correctness of current classification decisions in relation to past ones. We next examine this issue, as it presents fundamental contradictions in the formulation of the fairness-accuracy trade-off problem.

\subsubsection{Ignoring the Past}

As discussed in Section \ref{sec:intro}, optimization involves minimizing a loss function. In statistical terminology, minimizing the expected loss depends on the \emph{true} class labels \cite{bishop1995mlbook}. In fairness-related application domains we rarely, if ever, have access to true class labels. To return to an earlier example, Black people have systematically been denied loans in the US due to their race. In many cases, while a Black person's ``true" label should be that they would not default on a loan, past loan-granting decisions (and therefore the corresponding data) mark them as a defaulter. This captures the problem of \emph{label bias}: Misalignment between the ``ground truth" label and the actual, observed label in real world data (Section \ref{sec:preliminaries}). In a sense, this bias is what has motivated the entire field of algorithmic fairness in the first place: Automated decision systems that do not account for systemic discrimination in training data end up magnifying that discrimination \cite{barocas2018book, abebe2020social}; to avoid this, such systems need to be proactive about being fair.

Label bias presents an inherent issue with how we measure accuracy: If labels are wrong, particularly for individuals in the groups for which we want to increase fairer outcomes, then there are cases where misclassification is in fact the correct thing to do. In other words, how we measure accuracy is not truly accurate.

Yet, in the fairness-accuracy trade-off literature, it is very common to assume label bias can be ignored. Much of the work in this space does not mention label bias at all, or claims that it is out of scope for the research problem under consideration \cite{chen2018active}. This presents a contradiction: Simultaneously acknowledging that labels in the observed space are noisy representations of the ground truth (i.e., there is bias in the labels), but then explicitly assuming those labels in the training data (i.e., the observed space labels) are the same as the true labels \cite{dutta2020tradeoff}.\footnote{\citet{wick2019revisited} is a notable exception, acknowledging this contradiction in stark terms: ``...there is a pernicious modeling-evaluating dualism bedeviling fair machine learning in which phenomena such as label bias are appropriately acknowledged as a source of unfairness when designing fair models, only to be tacitly abandoned when evaluating them."} In other words, because the labels are biased, the corresponding accuracy measurements that depend on them are also biased; unfairness from the past produces inaccuracy \cite{hellman2020injustice}, which this work explicitly ignores in its trade-off formulation.

If accuracy measurements are conditioned on past unfairness, what is the trade-off between fairness and accuracy actually measuring? What does it mean to ``increase" or ``decrease accuracy" in this context? If accuracy measurements encode past unfairness for unprivileged groups, the fairness-accuracy trade-off is effectively positioning fairness in trade-off with unfairness, which is tautological. Giving validity to an accuracy metric that has a dependency on past unfairness inherently advantages privileged groups; it is aligned with maintaining the status quo, as there is no way to splice out the past unfairness on which it is conditioned. In the words of \citet{hellman2019fairness}, this can lead to even more unfair outcomes via ``compounding injustice." To the best of our knowledge, no prior work has explicitly attempted to avoid this scenario---to model and extract this past unfairness.%\footnote{No scholarship in this space, to the best of our knowledge, has attempted to put a Bayesian prior on existing unfairness. Such an approach would explicitly assume and model the existing unfairness due to a history of discrimination against certain demographics. Such modeling would not come without concern, as it would require fairness researchers to introduce a different set of mathematical modeling assumptions that carry their own normative implications.}

\subsubsection{Being Blind to the Future}

Similarly, studying specific, local classification decisions in terms of balancing fairness and accuracy does not provide insight about the more global, long-term effects that such decisions potentially have. This also presents a contradiction: Some scholarship concerning the trade-off explicitly aims to support the goals of policymakers, but policymaking by its very nature takes a long-tailed view. Current policy interventions do not just have a local impact, but rather also have desired cascading effects that carry into the future.

Ironically, this contradiction is clearly spelled out in some of the trade-off literature as an intentional assumption. For example, \citet{corbettdavies2017cost} explicitly states: ``Among the rules that satisfy a chosen fairness criterion, we assume policymakers would prefer the one that maximizes immediate utility." They intentionally examine the ``proximate costs and benefits" of the fairness-accuracy trade-off, assuming that this is the temporal resolution that would be most useful to policymakers. This approach enables simplifying mathematical assumptions, as it does not require evaluating how the specific automated decision under consideration has potential ramifications in the future. In \citet{corbettdavies2017cost}, in which they examine risk-assessment decisions for granting bail, they specifically do not need to look at how the immediate decision to detain someone may in fact be predictive of (even causally linked to) future arrests.  However, if such decisions are applied unfairly across racial demographic groups (even if somewhere slightly ``fairer" on the optimization curve), then they would just repeat patterns of bias existing in past data.

\subsection{Unfairness of ``Active Fairness" Trade-Off Remedies}

Some work regarding the fairness-accuracy trade-off sometimes goes beyond observing, characterizing, or implementing the trade-off for different applications, as we have discussed above. They note that while they agree with the notion that the trade-off is inherent, its effects can perhaps be mitigated by increasing \emph{both} accuracy and fairness---essentially, moving the trade-off optimization curve up and to the right (Figure \ref{fig:optimization}). The trade-off still exists in this scenario, but perhaps is less of an issue since the models perform better overall in terms of both accuracy and fairness.

Concretely, authors recommend a technique they call \emph{active feature acquisition} or \emph{active fairness} \cite{noriegacampero2019active}, which promotes the idea that ``data collection is often a means to reduce discrimination without sacrificing accuracy"  \cite{chen2018active}---that collecting more features \emph{for the unprivileged group} will help ensure fairer outcomes \cite{dutta2020tradeoff,bakker2019active}. The rationale is that additional feature collection alleviates the bias in the existing data for unprivileged groups, which will result in reduced bias in the classification results for those groups. Moreover, the authors note that gathering more features for the unprivileged group leads to these benefits without impacting the privileged group; the privileged group's accuracy and fairness metrics remain unchanged.

Setting aside that it might not even be possible to collect more features in practice, there are important implicit assumptions in this choice of solution. In particular, it seems like this work poses data collection as a value-neutral solution to ensure greater fairness. This assumption is clearly false. It is widely accepted, particularly in sociotechnical literature, that data collection is often a form of surveillance \cite{zuboff2018capitalism,clarke1994surveillance,brayne2017surveillance,cohen2011surveillance}. It is not a neutral act, and is generally not equally applied across demographic groups in the US. 

The choice to collect more data raises a normative question directly in contradiction with their goal for increased fairness for unprivileged groups: Do we really want to collect more data on unprivileged groups---groups that already tend to be surveilled at higher rates than those with more privilege? In the US in particular there is a long history of tracking non-white and queer individuals: Black Americans, from Martin Luther King to Black Lives Matter activists; Japanese Americans during World War II; non-white Muslims, particularly since 9/11; Latine individuals in relation to immigration status; trans, particularly trans-feminine, people in sports and bathroom use \cite{bedoya2016surveillance,desilver2020surveillance,speri2019surveillance,painter2011history, conrad2009trans, fl2021transbill}. In a more global treatment of fairness, is it fair to collect more data on these populations just to ensure we are optimizing some local fairness metric?

One could make the argument that machine learning broadly speaking pushes toward greater surveillance. The field is pushing to train larger and larger model specifications, which tend to require training on larger and larger datasets \cite{kaplan2020scaling}. These data-hungry methods in turn push for greater data collection and surveillance in general \cite{zuboff2018capitalism}. Yet, the proposed techniques in active fairness to alleviate the fairness-accuracy trade-off stand apart: They \emph{specifically} advocate for increasing data collection of already-surveilled groups. They tend to leave the data for the privileged group untouched in order to decrease classification disparities between groups. In essence, this unfairly shifts the burden of producing fair classification results to the unprivileged group, affording the additional privilege (i.e. even less relative surveillance) to the already privileged group. Put another way, their solution to the fairness-accuracy optimization problem introduces another, unexplored objective function---an objective function concerning the burden of surveillance, whose solution in this case causes residual unfairness for the marginalized group.

Some work in active fairness does acknowledge that additional data collection is not costless; however, this work often discusses it as a necessary cost for increased fairness rather than investigating it as a potential source for increased unfairness \cite{chen2018active}. \citet{noriegacampero2019active} states that it would be useful to model the cost of each feature in the dataset, where costs implicate monetary, privacy, and opportunity concerns. Beyond noting this idea, they do not attempt to formalize it in their work. 

\citet{bakker2019active} goes a step further by including cost in their problem formulation, associating a vector of costs with each feature. However, it is unclear how they pick the values of those costs and, perhaps more importantly, they make the assumption that the vector of costs is the same for each individual in the population. This assumes that different values for different features do not not incur different social impacts, which is demonstrably not the case. For example, consider individuals of transgender identity: Trans people, in comparison to cis people, face significant discrimination in response to disclosing their identity. In the language of \citet{bakker2019active}, it is more costly to trans people to collect features about gender identity than it is for cis people. In fact, such disparate costs can be thought of as the basis for needing to acknowledge and do policymaking using protected demographic attributes in the first place.

\section{Toward a Resolution} \label{sec:resolution}

\subsection{Making Normative Assumptions Explicit}

Writing mathematical proofs requires assumptions. For example, in machine learning, when writing proofs about an algorithm's properties, it is common to assume that the distribution we are trying to learn is convex. Assumptions like this enable us to guarantee certain logical conclusions about an algorithm's behavior, such as bounds on its convergence rate. While fairness-accuracy trade-off researchers are accustomed to stating mathematical assumptions like this, and ensuring that sound mathematical conclusions follow, we have shown that they do not pay similar attention to normative assumptions and their ensuing contradictory conclusions. We contend that researchers should take the time to make explicit such assumptions underlying their work. Being rigorous and clear about normative assumptions enables them to be reviewed just as rigorously as mathematical assumptions. 

We do not suggest that making such assumptions explicit is a sufficient solution on its own. Nevertheless, it would still help facilitate greater scrutiny about the appropriateness of proposed algorithmic fairness solutions. For example, as we discussed in Section \ref{sec:sociotechnical}, this would allow for considering that fairness and accuracy could in fact be in accord \cite{hellman2019fairness, selbst2019abstraction}. ML researchers should engage the assistance of social scientists if they believe they lack the expertise to do this work independently. Moreover, this process should facilitate researchers being introspective about how their individual backgrounds might inform the assumptions they bring into their work. This would be one necessary (though not on its own sufficient) way to address critics of fairness research being dominated by white voices \cite{abdurahman2019response}.

Moreover, clarifying implicit normative assumptions could facilitate rethinking how we measure accuracy. As we note in Section \ref{sec:assumptions}, common accuracy metrics are tied to unfairness. In an attempt to decouple accuracy from unfairness, one could, for example, put a Bayesian prior on existing unfairness and try to correct for it. To the best of our knowledge, no algorithmic fairness scholarship has attempted to do this: to explicitly assume and model the existing unfairness due to a history of discrimination against certain demographics.\footnote{Such modeling, of course, would not come without concern, as it would require introducing a different set of mathematical modeling assumptions that carry their own normative implications.}

\subsection{Tweaking Normative Assumptions for Robustness}

Making normative assumptions explicit could also help facilitate more robust ML fairness research. When investigating algorithmic robustness, researchers are generally comfortable with relaxing or changing certain mathematical proof assumptions and reasoning out the resulting changes (or stasis) in algorithm behavior.  As the economist Edward \citet{leamer1983econ} notes:

\begin{quote}
...an inference is not believable if it is fragile, if it can be reversed by minor changes in assumptions. … A researcher has to decide which assumptions or which sets of alternative assumptions are worth reporting.
\end{quote}

As a test of normative robustness, we similarly recommend that fairness-accuracy trade-off researchers perturb their normative assumptions and investigate how this may alter normative outcomes. For example, when considering surveillance of the unprivileged via active feature acquisition as an appropriate mechanism for alleviating the trade-off, it would be useful to state this as an explicit assumption, and then consider surveillance as a constraint for the problem. In other words, one could ask, how much surveillance is tolerable for increased fairness? Perhaps none, but perhaps there is a small set of high quality features that could be collected to serve this purpose, rather than just indiscriminately collecting a lot of additional features.

\section{Conclusion: Reconsidering the Fairness-Accuracy Trade-off} \label{sec:conclusion}
As \citet{passi2019formulation} note, ``Whether we consider a data science project fair often has as much to do with the formulation of the problem as any property of the resulting model." Furthermore, the work of problem formation is ``rarely worked out with explicit normative considerations in mind" \cite{passi2019formulation}. As we have shown in this article, not attending explicitly to these considerations can lead to contradictory, unintended results: Formulating a trade-off between fairness and accuracy involves a variety of normative assumptions that can in fact lead to various forms of emergent unfairness. 

Our recommendations in Section \ref{sec:resolution} to make normative assumptions explicit aim to remedy this particular type of unfair outcome. However, we recognize that in the case of the fairness-accuracy trade-off, fully applying these recommendations may not be sufficient in itself. Rather, in clarifying the normative concerns of the trade-off, it is quite possible to reasonably conclude that the effects of emergent unfairness outweigh any benefits that come from choosing this particular problem formulation. As such, when it comes to the critical issue of algorithmic fairness, it may be time to reconsider the framing of trade-offs altogether.

%Passi: Problem formulation as part of the issue with fairness research. “Our research demonstrates that the specification and operationalization of the problem are always negotiated and elastic, and rarely worked out with explicit normative considerations in mind.

%\citet{gieren1983boundary} has famously discussed this legitimacy in terms of ``boundary-work": Scientists have an ideological style in which they cordon off science into its own sphere, which they frame as superior to non-scientific intellectual practices.

\begin{acks}
Thank you to our anonymous reviewers for their valuable feedback. Additionally, we would like to thank the following individuals for feedback on earlier drafts and iterations of this work: Professor Rediet Abebe, Bilan A.H. Ali, Harry Auster, Kate Donahue, Professor Chris De Sa, Jessica Zosa Forde, Professor Deborah Hellman, and Kweku Kwegyir-Aggrey. We would also like to thank the Artificial Intelligence Policy and Practice initiative at Cornell University, the John D. and Catherine T. MacArthur Foundation, and the Cornell Humanities Scholars Program.
\end{acks}

\bibliographystyle{ACM-Reference-Format}
\balance
\bibliography{references}

\end{document}